\documentclass[a4paper]{article}

\usepackage{INTERSPEECH2020}

\title{Cross-task pre-training for on-device acoustic scene classification}
\name{Ruixiong Zhang, Wei Zou, Xiangang Li}
\address{Didi Chuxing, Beijing, China}
\email{\{zhangruixiong, zouwei, lixiangang\}@didiglobal.com}

\begin{document}

\maketitle
\begin{abstract}
Acoustic scene classification (ASC) and acoustic event detection (AED) are different but related tasks. Acoustic events can provide useful information for recognizing acoustic scenes. However, most of the datasets are provided without either the acoustic event or scene labels. To utilize the acoustic event information to improve the performance of ASC tasks, we present the cross-task pre-training mechanism which utilizes acoustic event information from the pre-trained AED model for ASC tasks. On the other hand, most of the models were designed and implemented on platforms with rich computing resources, and the on-device applications were limited. To solve this problem, we use model distillation method to compress our cross-task model to enable on-device acoustic scene classification. In this paper, the cross-task models and their student model were trained and evaluated on two datasets: TAU Urban Acoustic Scenes 2019 dataset and TUT Acoustic Scenes 2017 dataset. Results have shown that cross-task pre-training mechanism can significantly improve the performance of ASC tasks. The performance of our best model improved relatively 9.5\% in the TAU Urban Acoustic Scenes 2019 dataset, and also improved 10\% in the TUT Acoustic Scenes 2017 dataset compared with the official baseline. At the same time, the performance of the student model is much better than that of the model without teachers.
\end{abstract}

\section{Introduction}
\label{sec:intro}
ASC and AED are different but related tasks\cite{schuller2018interspeech,schuller2019interspeech,mesaros2017dcase,tzanetakis2000marsyas}. For ASC, we consider that acoustic events can provide additional useful information because human listeners can easily recognize acoustic scenes based on acoustic events\cite{schroeder2011detection,guo2017attention,tonami2019joint}. Table 1 shows three acoustic scenes with their corresponding acoustic events. We can see that the acoustic scenes can be easily identified given the corresponding acoustic events. It is expected that the acoustic event information can be utilized to further improve the performance of the ASC model.

To accomplish this task, both acoustic event and scene labels of the dataset are needed. However, it would be harder to collect data and the data size would also be limited. Meanwhile, most of the open source datasets are provided without either the acoustic event or scene labels\cite{schuller2018interspeech,schuller2019interspeech,mesaros2016tut,giannoulis2013database}. In this paper, we introduce pre-training mechanism for cross-task learning. Specifically, we firstly pre-train a model for AED using the dataset with only acoustic event labels. Secondly we utilize the acoustic event representations from the pre-trained model to train a new model for ASC using the target dataset with only acoustic scene labels. To use pre-trained representations to downstream tasks, there are two strategies according to \cite{devlin2018bert}: feature-based and fine-tuning. We analyzed the performance of these two strategies in cross-task learning. To our knowledge, it is the first time to utilize acoustic event information in a pre-trained manner to improve the performance of the ASC model. 

\begin{table}[t]
	\caption{Acoustic scenes and their corresponding acoustic events}
	\label{tab:Recording examples}
	\setlength{\tabcolsep}{5pt} 
	\centering
	\begin{tabular}{ll}
		\toprule
		\textbf{Acoustic Events} & \textbf{Acoustic Scenes} \\
		\midrule
		traffic noise, roadway noise, crowed noise & Street with traffic \\ 
		report, aeroengine noise, crowed noise & Airport \\
		music, conversation, TV sound & Home \\ 
		\bottomrule
	\end{tabular}
\end{table}

To achieve the best performance, large models with tens of millions of parameters would be built. However, it is not applicable to the mobile devices with limited computation and storage resources\cite{fu2019mobile, fukuda2017efficient}. To solve this problem, we compress our cross-task models to enable on-device detection. We use knowledge distillation\cite{kim2016sequence, chen2017learning} method to compress our cross-task model with slight accuracy decline.

\section{Related Work}
\label{sec:Related Work}

\cite{nguyen2019acoustic} has proposed a heterogeneous system called Deep Mixture of Experts (DMoEs). Each DMoEs module is a mixture of different parallel CNN structures and these modules are weighted by a gating network. This system also utilizes pre-training mechanism in which all CNNs are firstly pre-trained to be used to extract a variety of features. After that, the CNNs would be fixed and the gating network would be updated to estimate the contribution of each module in the mixture. To explore the performance, they used an ensemble of three DMoEs modules each with different pairs of inputs and individual CNN models. The input pairs are spectrogram combinations of binaural audio and mono audio as well as their pre-processed variations using harmonic-percussive source separation (HPSS) and nearest neighbor filters (NNFs). The experimental result of this system is 72.1\% improving the baseline by around 12\% on the development data of DCASE 2018 task 1A\cite{Mesaros2018_DCASE}. 

\cite{tonami2019joint} has proposed multitask learning for joint analysis of acoustic events and scenes. It utilized acoustic scene information to improve the performance of AED. Experimental results obtained using TUT Sound Events 2016/2017\cite{Mesaros2019_TASLP} and TUT Acoustic Scenes 2016 datasets\cite{Mesaros2018_TASLP} have shown that this proposed method can significantly improve the performance of acoustic event detection compared with the baseline. Different from this work, we utilized acoustic event information in a pre-trained manner to improve the performance of ASC. 

\section{Methodology}
\label{sec:methodology}
For the cross-task model, it is composed of two training stages. In the pre-training stage we use VGGish\cite{gemmeke2017audio} to build an AED model. In the next stage, the pre-trained VGGish would be utilized to build an ASC model. The trained ASC model would be then utilized as the teacher to train a smaller model which is called the student. To train the student model, knowledge distillation method would be utilized. The details would be discussed below.

\subsection{Cross-task models}
\label{sec:Teacher}
Our base model is composed of a simple CNN based encoder and an attention based pooling layer. To combine the pre-trained VGGish, we learn from previous studies\cite{vaswani2017attention,devlin2018bert,baltruvsaitis2018multimodal} and propose three cross-task models: VGGish Feature Based Model (VFM) which directly replace the raw feature with pre-trained VGGish features, Joint Representation based Model (JRM) which treats the raw feature and the pre-trained feature as two modals, Cross-task Multi-head Attention Model (CMAM) which utilizes multi-head attention mechanism to combine information from different sources. All of them are built on the same base model for comparison.
\subsubsection{Base}
\label{sec:Base}
Because of the previous successful application of CNN-based layers to audio classification tasks\cite{hershey2017cnn}, our base model uses CNN-based layers to extract high level features. Its structure is shown in Figure 1.
\begin{figure}[htb]
	\begin{minipage}[b]{1.0\linewidth}
		\centering
		\centerline{\includegraphics[width=3.5cm,height=4.5cm]{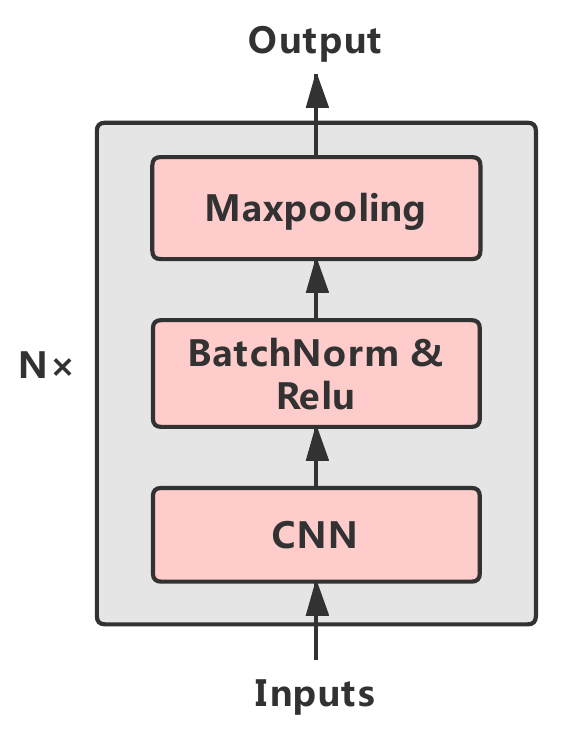}}
		\caption{Structure of CNN-based encoder}\medskip
	\end{minipage}
\end{figure}

The pooling layer is used to map temporal representations to one representation\cite{45611,barchiesi2015acoustic}. The attention-based pooling layer with multiple heads outperforms that with a single head and has achieved superior performance in the area of speaker recognition\cite{zhu2018self}, so we use it as our pooling layer. Its structure is shown in Figure 2.
\begin{figure}[htb]
	\begin{minipage}[b]{1.0\linewidth}
		\centering
		\centerline{\includegraphics[width=3cm,height=3.5cm]{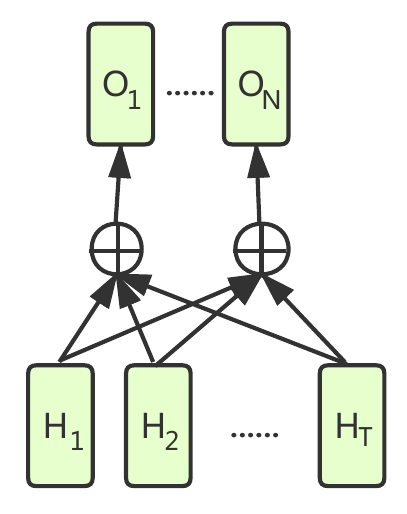}}
		\caption{Structure of attention-based pooling function with multiple heads}\medskip
	\end{minipage}
\end{figure}
\subsubsection{VGGish Feature based Model}
\label{sec:vggish}
In most unsupervised pre-training cases, the raw features are replaced with pre-trained features\cite{erhan2010does}. We get experimental experiences from unsupervised pre-training models and explore the performance of this simple strategy in the cross-task case. As shown in Figure 3, the main difference between the base model and VFM is the input to the CNNs. 
\begin{figure}[htb]
	\begin{minipage}[b]{1.0\linewidth}
		\centering
		\centerline{\includegraphics[height=5cm]{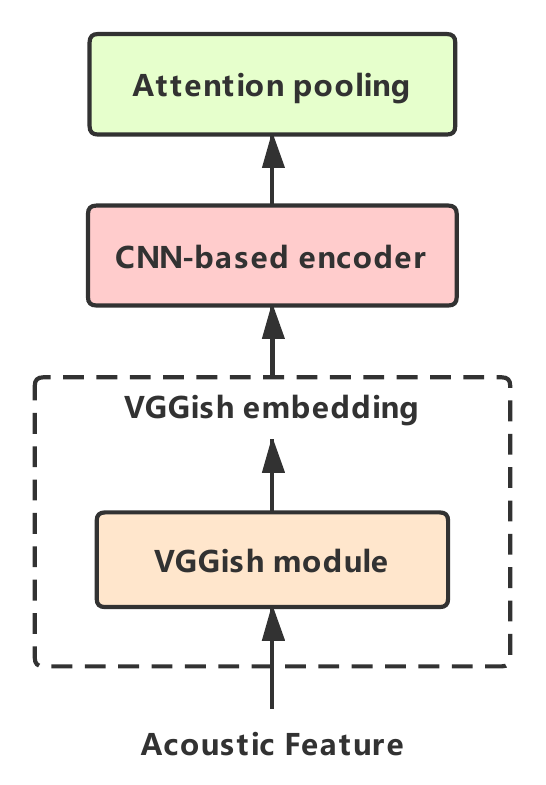}}
		\caption{Structure of VFM}\medskip
	\end{minipage}
\end{figure}
\subsubsection{Joint Representation based Model}
\label{sec:joint}
Inspired by \cite{baltruvsaitis2018multimodal}, the pre-trained VGGish feature contains acoustic event information part of which can never be learned from raw features. To maximize the information usage from the pre-trained VGGish feature, we can treat the raw feature and the pre-trained VGGish feature as two separate representations. As shown in Figure 4, after a series of separate calculations, the representations from the raw feature and the pre-trained VGGish feature would be concatenated. The concatenated representation is then fed into the fully-connected layer to predict the probabilities of target labels.
\begin{figure}[htb]
	\begin{minipage}[b]{1.0\linewidth}
		\centering
		\centerline{\includegraphics[height=5cm]{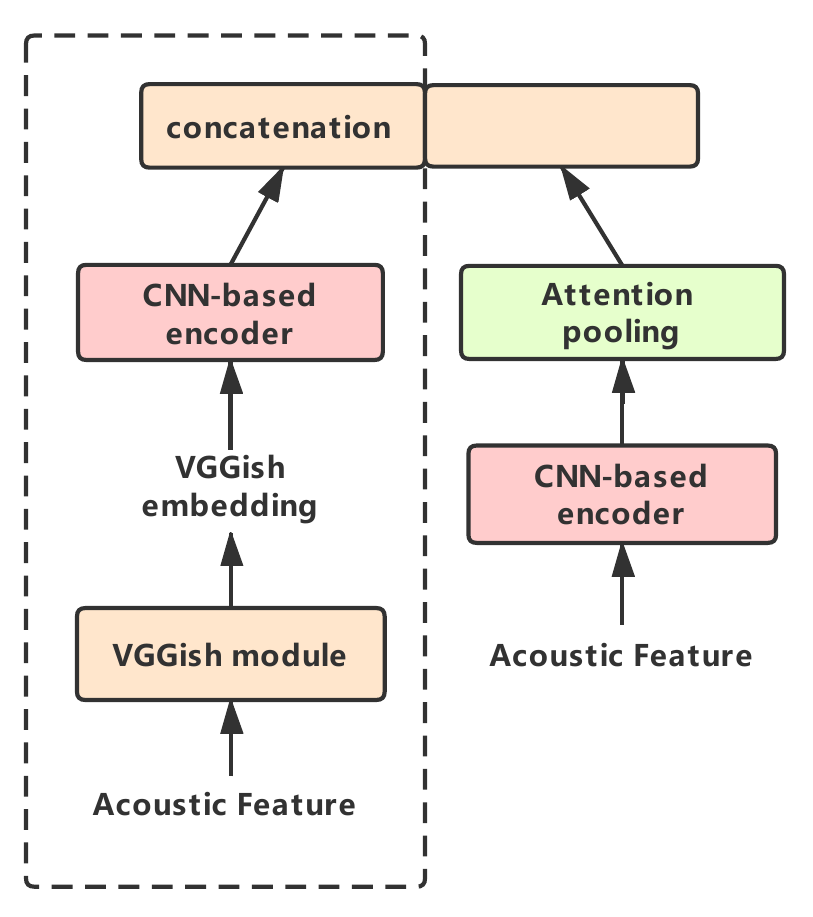}}
		\caption{Structure of JRM}\medskip
	\end{minipage}
\end{figure}
\subsubsection{Cross-task Multi-head Attention Model}
\label{sec:mult-head}
For JRM, the high level features from the raw feature and the pre-trained VGGish feature are hard concatenated. It means that the contribution of the pre-trained feature can still not be adjusted properly. On the other hand, transformer-based models have achieved great performance in many kinds of acoustic fields\cite{zhou2018syllable,lian2018improving}. Therefore, CMAM is presented. Unlike the hard concatenation of representations in JRM, the contribution of the VGGish feature in CMAM is more dynamic. 
\begin{figure}[htb]
	\begin{minipage}[b]{1.0\linewidth}
		\centering
		\centerline{\includegraphics[height=6cm]{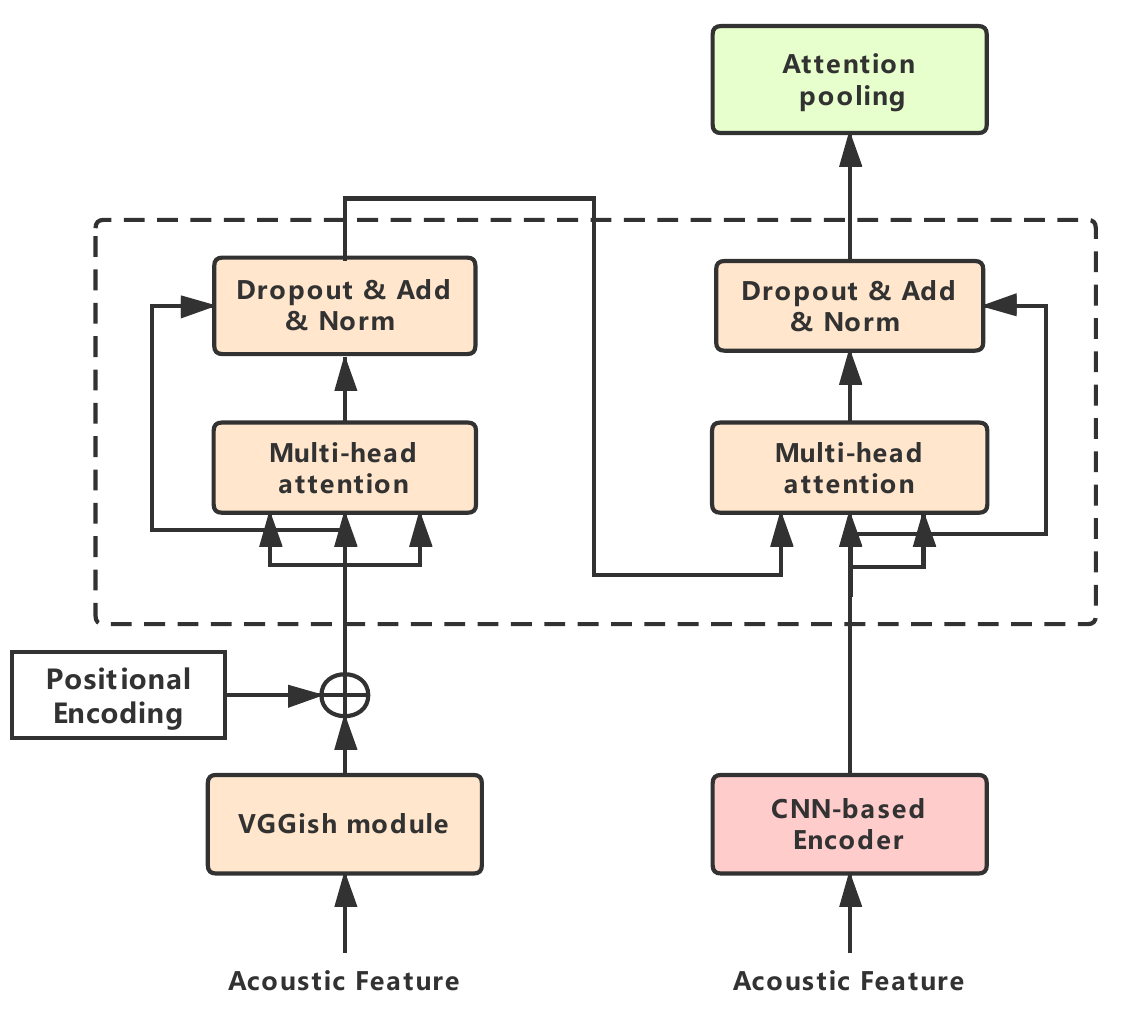}}
		\caption{Structure of CMAM}\medskip
	\end{minipage}
\end{figure}
As shown in Figure 5, we firstly inject positional information as in \cite{vaswani2017attention} into the VGGish representation. And then the VGGish representation is fed into the multi-head self attention layer and computed as follow equations:
\begin{gather}
	Q = HW^Q \\
	K = HW^K\\
	V = HW^V
\end{gather}
where $H$ is the VGGish representation and $W^Q, W^K$ and $W^V$ are learnable parameter matrices. The output $Q, K, V$ are then split into $N$ heads on the channel dimension and are computed as follow equations:
\begin{gather}
	O_i = softmax(\frac{Q_iK_i^T}{\sqrt{d}})V_i\\
	O_a = Concat(O_iW^O)\\
	O = LayerNorm(Dropout(O_a) + H)
\end{gather}
where $Q_i, K_i, V_i$ are the $i$-th head of $Q, K, V$ and the $Concat$ function concatenates outputs of all heads. $d$ is the dimension of queries. $W^O$ is the learnable parameters. $O$ is the final output of this layer.

$O$ is then used to attend features from the CNN-based encoder. The multi-head attention layer here is same as that mentioned above except that $Q$ is the output from VGGish and $K, V$ are the outputs from the CNN-based encoder.
\subsection{Model Distillation}
\label{sec:Model Distillation}
As mentioned above, we proposed three kinds of models (VFM, JRM, CMAM) for training. However, All of them  have large parameters and they are not applicable to the mobile devices with limited computation and storage resources. We use model distillation method to transfer the knowledge of these models to a smaller model.

Various algorithms have been proposed for model distillation\cite{sun2019patient,perez2020audio}. Instead of using ground truth labels, the loss function of teacher-student training can be defined as:
\begin{gather}
	L = -\sum_{i} q_i \log p_i
\end{gather}
where $q_i$ represents the soft labels from the teacher model, and $p_i$ represents the output probability of the class of the student model. Different from training using ground truth labels, the classes in $q_i$ will have small posterior probabilities for each training example instead of zeros. We also utilize the augmented-training method mentioned in \cite{fukuda2017efficient} to train the student model using two or three of them simultaneously.

For the student model, we use the base model mentioned above. The base model can be used for deployment because it only contains hundreds of thousands of parameters, which can reduce the storage usage and effectively speed the forward propagation time up.
\section{Experiments}
\label{sec:experiments}
\subsection{Dataset}
\label{sec:dataset}
For pre-training, The AudioSet dataset\cite{gemmeke2017audio} is utilized to train VGGish. It is a large-scale collection of human-labeled 10-second sound clips drawn from YouTube videos. It contains totally 2084320 videos which contain 527 labels.

For cross-task training, we use the TUT Acoustic Scenes 2017 development dataset\cite{mesaros2017dcase} and the TAU Urban Acoustic Scenes 2019 development dataset\cite{Mesaros2018_DCASE}. As mentioned in \cite{mesaros2017dcase}, the TUT Acoustic Scenes 2017 development dataset contains 15 acoustic scene labels and audio segments with a length of 10 seconds. Each acoustic scene has 312 segments totaling 52 minutes of audio. The experiments in TUT Acoustic Scenes 2017 dataset were conducted using a four-fold cross-validation setup. The TAU Urban Acoustic Scenes 2019 development dataset contains 10 acoustic scene labels and the length of audio is also 10 seconds. Each acoustic scene has 1440 segments totaling 240 minutes of audio. The dataset contains in total 40 hours of audio.

\subsection{Experimental setups}
All the models use log-mel energies of 64 dimensions extracted from 25ms segments with 10ms overlap as the acoustic feature. For the base model and VFM, the CNN-based encoder is composed of 4 CNN layers with the kernel size of 3 and the max pooling size of 2. The channel sizes of these CNN layers are 64, 128, 256, 512. For JRM, its encoder is composed of 7 CNN layers with the kernel size of 3 and the max pooling size of 2. The channel sizes of these CNN layers are 64, 128, 256, 512, 1024, 1024, 2048. For CMAM, the encoder is composed of 12 multi-head attention layers with the attention heads of 8 and the query dimension of 2048. All the models use the attention-based pooling layer with the attention heads of 4 and the attention dimension of 1024. We add 1 feed-forward layer after the attention-based pooling layer to project the hidden representations to the output probabilities. 

All the models were trained on one GPU with the batch size of 64 for 200 epochs. We used the Adam optimizer\cite{kingma2014adam} with the learning rate of 0.001 for the base model, VFM and JRM. For CMAM, we used the Adam optimizer with warmup schedule\cite{vaswani2017attention} according to the formula:
\begin{equation}
lrate = k * d_{model}^{0.5} * min(n^{-0.5}, n*warmup\_n^{-1.5})
\end{equation}
where $n$ is the step number. k = 0.5 and warmup\_n = 8000 were chosen for the experiment.

For the teacher models, we train them in two strategies respectively: the feature-based strategy in which the pre-trained VGGish parameters are fixed, the fine-tuning strategy in which the pre-trained VGGish parameters are fine-tuned. For comparison, we also train the whole model from scratch. For the student model, the structure is same as the base model. We use different teacher models and their combinations to train the student model.

We use classification accuracy as our metrics. It is calculated as average of the class-wise accuracy which is calculated as the number of correctly classified segments among the total segment number of this class. 

\subsection{Experimental Results And Discussion}
\label{sec:Conclusion}
\subsubsection{Cross-task models}
\label{sec:Teacher models}
Results of teacher models are shown in Table \ref{tab:Experimental results of teacher models}. For VFM, we can see that VFM performed the worst. It is reasonable because unlike unsupervised pre-training in which the model is trained using fundamental information like the co-occurrence of words in sentences in NLP tasks\cite{devlin2018bert}, cross-task pre-training leads to inevitable mix of unrelated noise with the useful information. On the other hand, unlike JRM with a relative deep CNN-based encoder to process the pre-trained feature, VFM has no such structure to support cross-task learning. 

For JRM, we can see that in the feature-based strategy, unstable performance was presented. The performance is better than that of the model trained from scratch in the TUT Acoustic Scenes 2017 development dataset. However, the performance is worse than that of the model trained from scratch in another dataset. In the fine-tuning strategy, the performance is also unstable. This is mainly due to the hard concatenation of the high level features from the raw feature and the pre-trained VGGish feature. The contribution of the pre-trained feature can still not be adjusted properly. 

For CMAM, on one hand, the feature-based strategy performed better than the strategy from scratch. It indicated that CMAM has effectively utilized cross-task pre-trained information. On the other hand, the performance is much better than that of JRM although the number of parameters in CMAM and JRM are similar. 
We considered that multi-head attention mechanism in CMAM has more advantages to dynamically adjust the contribution of the pre-trained feature. We can also see that CMAM with the feature-based strategy performed the better than that with the fine-tuning strategy. The performance in the fine-tuning strategy is similar with that of the model trained from scratch. We considered that this is mainly because we use the same learning rate in the whole model. In this situation, there is no proper global learning rate to converge the model in a suitable location and guarantee that the pre-trained information can not be washed away.

\begin{table}[t]
	\caption{Experimental results of teacher models (Note: TUT represents the TUT Acoustic Scenes 2017 dataset and TAU represents the TAU Urban Acoustic Scenes 2019 dataset)}
	\label{tab:Experimental results of teacher models}
	\setlength{\tabcolsep}{10pt} 
	\centering
	\begin{tabular}{lll}
		\toprule
		Model Type & \textbf{TUT(\%)} & \textbf{TAU(\%)} \\
		\midrule
		Official base\cite{mesaros2017dcase, Mesaros2018_DCASE} & 78.4 & 62.5 \\ 
		\midrule
		Our base & 81.7 & 66.5 \\
		\midrule
		VFM from scratch & - & - \\
		VFM feature based & 72.1 & 57.3 \\
		VFM fine-tuning & - & - \\
		\midrule
		JRM from scratch & 83.0 & 67.2 \\
		JRM feature based & 83.7 & 63.4 \\
		JRM fine-tuning & 83.2 & 66.8 \\
		\midrule
		CMAM from scratch & 83.2 & 65.1 \\
		CMAM feature based & {\bfseries 85.8} & {\bfseries 68.8} \\
		CMAM fine-tuning & 83.1 & 65.3 \\		
		\bottomrule
	\end{tabular}
\end{table}
\subsubsection{model distillation}
\label{sec:model distillation}
For model distillation, we used our base model as the student model. We used JRM, CMAM as the teacher model and explored their performance which are shown in Table \ref{tab:Experimental results of model distillation}. We can see that JRM in both the feature-based strategy and the fine-tuning strategy can slightly improve the performance of the student and JRM in the fine-tuning strategy performed slightly better. Compared to JRM, CMAM performed much better and CMAM in the feature-based strategy significantly improved the performance of the student. It is correlated with the performance of the teacher model itself as discussed above. It also indicated that the acoustic event information from the teacher can be transferred to the student.

On the other hand, we use the augmented-training method to combine CMAM and JRM to train the student. We 
explored the performance of the ensemble of JRM in the fine-tuning strategy and CMAM in the feature-based strategy. We can see that the student trained by the ensemble of teachers performed better than any student trained by a single teacher. It indicated that different cross-task models can provide complementary information to improve the performance of the student.

\begin{table}[t]
	\caption{Experimental results of model distillation (Note: TUT represents the TUT Acoustic Scenes 2017 dataset and TAU represents the TAU Urban Acoustic Scenes 2019 dataset, JRM + CMAM represents the 
		ensemble of JRM in the fine-tuning strategy and CMAM in the feature-based strategy)}
	\label{tab:Experimental results of model distillation}
	\setlength{\tabcolsep}{10pt} 
	\centering
	\begin{tabular}{lll}
		\toprule
		Teacher Type & \textbf{TUT(\%)} & \textbf{TAU(\%)} \\
		\midrule
		- & 81.7 & 66.5 \\
		\midrule
		JRM (feature-based) & 82.0 & 66.6 \\
		JRM (fine-tuning) & 82.1 & 66.7 \\
		CMAM (feature-based) & 83.1 & 67.2 \\
		CMAM (fine-tuning) & 82.6 & 66.8 \\
		JRM + CMAM& {\bfseries 84.3} & {\bfseries 67.7} \\
		\bottomrule
	\end{tabular}
\end{table}

\section{Conclusion}
\label{sec:Discussions And Conclusion}
In this paper, we explored cross-task pre-training for on-device acoustic scene classification. We presented three cross-task models including VFM, JRM and CMAM and we explored feature-based and fine-tuning strategies. For on-device application, we use model distillation and use the cross-task model as the teacher model. Meanwhile, we use the augmented-training method to combine JRM and CMAM to train the student. Experimental results indicated that for the teacher model, JRM and CMAM both can utilize the acoustic event information from the pre-trained VGGish and CMAM in the feature-based strategy has more advantages in cross-task learning. For model distillation, both CMAM and JRM can effectively improve the performance of the student and the ensemble of JRM and CMAM performed the best. 

\bibliographystyle{IEEEtran}

\bibliography{refs}


\end{document}